\documentclass[pdflatex,sn-basic]{sn-jnl}

\usepackage{graphicx}%
\usepackage{multirow}%
\usepackage{amsmath,amssymb,amsfonts}%
\usepackage{amsthm}%
\usepackage{mathrsfs}%
\usepackage[title]{appendix}%
\usepackage{xcolor}%
\usepackage{textcomp}%
\usepackage{manyfoot}%
\usepackage{booktabs}%
\usepackage{algorithm}%
\usepackage{algorithmicx}%
\usepackage{algpseudocode}%
\usepackage{listings}%
\usepackage{amsmath}%
\usepackage{parskip}%
\usepackage{natbib}%
\usepackage{subcaption}
\usepackage{siunitx}
\usepackage{parskip}
\newcommand{\apj}{{\it Astrophys. J.}}
\newcommand{\apjl}{{\it Astrophys. J. Lett.}}
\newcommand{\solphys}{{\it Sol. Phys.}}
\newcommand{\aap}{{\it Astron. Astrophys.}}
\newcommand{\ssr}{{\it Space Sci Rev.}}

\newcommand{\mnras}{{\it Monthly Notices of Royal Astronomical Society.}}

\newcommand{\nat}{{\it Nature.}}

\newcommand{\apss}{{\it Astrophysics and Space Science.}}

\begin{document}

\title[Article Title]{Spectroscopic Diagnosis of a B-Class Flare and an Associated Filament Eruption}


\author[1]{\fnm{B. Suresh Babu}}
\email{sureshsuju96@gmail.com}

\author*[1]{\fnm{Pradeep} \sur{Kayshap}}\email{virat.com@gmail.com}
\equalcont{These authors contributed equally to this work.}

\author[1]{\fnm{Sharad C} \sur{Tripathi}}\email{risingsharad@gmail.com}
\equalcont{These authors contributed equally to this work.}

\affil*[1]{\orgdiv{School of Advanced Sciences and Languages}, \orgname{VIT Bhopal University}, \orgaddress{\street{Kothri-Kalan}, \city{Sehore}, \postcode{466114}, \state{M.P}, \country{India}}}


\abstract{The flare ribbon and an associated filament eruption are diagnosed using O~{\sc iv}~1401.16~{\AA}, Si~{\sc iv}~1402.77~{\AA}, and Mg~{\sc ii} k 2796.35~{\AA} spectral lines provided by IRIS. The flare ribbons have downflow (redshifts) in all these lines, and this redshift decreases from the transition region to the chromosphere. While the overlapping region (flare-ribbon+filament rise/eruption is dominated by upflows (blueshifts) in all three spectral lines. We found an extremely blueshifted Si~{\sc iv} profile (i.e., blueshift around -180 km/s) in the overlapping region. The {mean non-thermal velocity (v$_{nt}$)} in the flare ribbons is higher in O~{\sc iv} than Si~{\sc iv}. While, in the overlapping region, O~{\sc iv} have lower v$_{nt}$ than Si~{\sc iv}. Note that very high v$_{nt}$ around 80 km/s (in Si~{\sc iv}) exists in this weak B-class flare. The Mg~{\sc ii} k line widths are almost the same in the flare ribbon and overlapping region but, they are extremely broad than previously reported. We found double peak profiles of Si~{\sc iv} and O~{\sc iv} in the overlapping region. Most probably, one peak is due to downflow (flare ribbon) and another due to upflow (filament rise/eruption). We report a high redshift of more than 150 km/s in the weak B-class flare. In some cases, both peaks show upflows which might be the result of the superposition of two different sources, i.e., overlapping of two different velocity distributions in the line of sight.}




\maketitle

\section{Introduction}\label{sec:intro}
Mostly the solar flares occur in the active regions (ARs), and they are very energetic phenomena of the solar atmosphere (\cite{2017LRSP...14....2B}). The magnetic energy converts into radiative, thermal, and kinetic energy during the solar flares via the magnetic reconnection process (\cite{2011LRSP....8....6S, 2017LRSP...14....2B}). Some of the solar flares are associated with coronal mass ejection (CMEs), known as eruptive flares. Meanwhile, the flares with no CMEs are known as confined solar flares. According to the CSHKP model, the reconnection takes place in the solar corona, and high-energy particles precipitate in the lower solar atmosphere, and form the flare ribbons (e.g., \cite{1964NASSP..50..451C, 1966Natur.211..695S, 1974SoPh...34..323H, 1976SoPh...50...85K}).\\
Spectroscopic diagnosis is an important tool for gathering information about heat and energy transfer over any particular region of the solar atmosphere, for instance, flare ribbons. The spectroscopic diagnosis helps us to deduce the intensity, plasma flow, non-thermal motions, and electron density from a region of interest (e.g., \cite{2018LRSP...15....5D}). Next, we mention that the strength and direction of the plasma flow can be inferred from the Doppler velocities forming over the different heights of the solar atmosphere (e.g., \cite{2018LRSP...15....5D}). The flare ribbons have upflows in the coronal lines which is the result of the chromospheric evaporation (e.g., \cite{2015ApJ...799..218Y, 2017LRSP...14....2B}). While, cool lines forming in the transition region (TR) and chromosphere are dominated by the downflows resulting from the chromospheric condensation (e.g., \cite{1995SoPh..158...81D, Brosius2015, K_Yu2020}). Both physical processes (i.e., chromospheric evaporation and condensation) are the secondary process of the magnetic reconnection that triggers the solar flares (e.g., \cite{2011ApJ...731L...3S}).\\
Apart from the Doppler velocity (v$_{d}$), another spectroscopic parameter v$_{nt}$ tells about the unresolved motions in any particular plasma volume (e.g., \cite{1993SoPh..144..217D, 1997SoPh..173..243D, 2023MNRAS.526..383K, 2024ApJ...977..141K}). In the extreme ultraviolet (EUV) and soft X-ray lines, the v$_{nt}$ can increase up to 200 km/s during the solar flares (\cite{1980ApJ...239..725D}). While, in the case of the ultraviolet line (i.e., Si~{\sc iv}), the v$_{nt}$ ranges 40{--}60 km/s in the solar flare (\cite{2015ApJ...809...46C}). Not only during the flare but the enhancement in the nonthermal velocity appears before the flare too (i.e., pre-flare phase; \cite{2013ApJ...774..122H}). In the solar flares, the unresolved turbulent motions in the emitting region or superposition of multiple sources along the line-of-sight are the most probable reasons behind the excessive broadening of the spectral lines (e.g., \cite{2011ApJ...740...70M, 2018ApJ...854..122W, 2024ApJ...975...33C}). Significant advances in the flare diagnostic have been made using spectroscopic observations still, our understanding of the solar flares (i.e., nature of profiles, v$_{d}$ structure, and nature of nonthermal broadening) is very weak and needs further investigations.\\

This work is focused on the spectroscopic diagnosis of the flare ribbons. In this event, just after the solar flare, one filament also erupted and overlapped with the lower part of the flare ribbon. Section~\ref{sect:obs} describes the observations and data analysis. The observational results are described in Section~\ref{sect:res}. Finally, the discussion and conclusions are described in the last section.

\section{Observations and Data Analysis}\label{sect:obs}
We have diagnosed the flare-ribbon and an associated filament eruption. The event has occurred from an active region (AR) 12661 on June 07$^{th}$, 2017 from 14:00$-$19:00~UT, and it was observed by Interface Region Imaging Spectrograph (IRIS). IRIS has observed the spectra of this event in the dense raster mode, having 192 steps with the slit width of 0.35$"$. The pixel size along the slit is 0.16$"$, and there are total 773 pixels along the slit. Therefore, the field of view (FOV) covered in this observation is 67$"$x128$"$, i.e., 192 (steps)$\times$0.35$"$ = 67$"$ and 773 (pixels in Y)$\times$0.166$"$ = 128$"$. Finally, we mention that the step cadence is 5 s. IRIS has captured the spectra in the far-ultraviolet (FUV) and near-ultraviolet (NUV) waveband of the solar spectrum, and these wavebands include several absorption and emission lines. See \cite{Bart_2014} for more details on the IRIS spectra. The Si~{\sc iv} 1402.77 {\AA} (log T = 4.8~K), O~{\sc iv} 1401.16~{\AA} (log T = 5.2~K), and Mg~{\sc ii k} 2796.20 {\AA} (log T = 4.0~K) are used to diagnose the flare/filament plasma in transition-region and chromosphere.\\

The Si~{\sc iv} 1402.77~{\AA} and O~{\sc iv} 1401.16~{\AA} lines are optically thin lines. Therefore, we fit these lines using a single Gaussian to deduce peak intensity, centroid, and Gaussian width at each location of the observed region. Next, we estimate the total intensity (I$_{tot}$) and the v$_{nt}$ (estimated using iris$\_$nonthermalwidth.pro routine). Lastly, the centroid is converted to the v$_{d}$ using the rest wavelength. The rest wavelengths of Si~{\sc iv} and O~{\sc iv} are estimated using the cool line (i.e., C~{\sc i} 1404.29~{\AA}) method (e.g., \cite{1999ApJ...522.1148P, 2024aApSS...369...61B, 2024bMNRAS...528...2474B}). Mg~{\sc ii k} 2796.35~{\AA} line is an optically thick line that forms in the chromosphere. It shows two peaks (k2v and k2r) and a central reversal (k3) in the quite-Sun (QS) (e.g., \cite{2013ApJ...772...90L, 2018ApJ...864...21K}), but it mostly appears as a single peak profile in the flare ribbons (e.g., \cite{kerr2015}). However, this is not always the case, the Mg~{\sc ii} k line from the flare ribbons might show two peaks with the central reversal similar to QS. Therefore, we adopted a non-parametric approach (quartile method; \cite{kerr2015}), instead of the Gaussian fit, to deduce the v$_{d}$, line width, and asymmetry.
\begin{figure*}
\centering
\includegraphics[trim=4.0cm -0.4cm 3.0cm 0.0cm, scale=0.8]{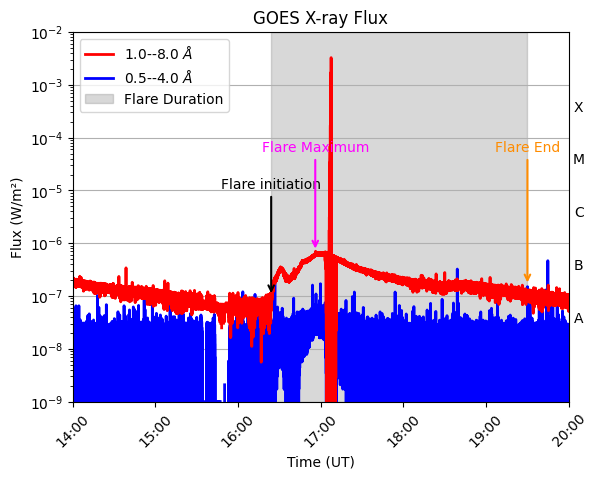}
\caption{The figure shows GOES X-ray flux from 14:00 to 20:00~UT on June 07$^{th}$, 2017. The red curve shows the soft X-ray (i.e., 1.0{--}8.0~{\AA}, and the blue curve shows the hard X-ray (i.e., 0.5{--}4.0~{\AA}). The flare's total duration is more than 3 hours, and it is highlighted by the grey area. The flare starts at 16:24~UT (indicated by a black arrow), peaks at 16:55~UT (indicated by a magenta arrow), and completes around 19:30~UT (indicated by an orange arrow).}
\label{fig:fig_goes}
\end{figure*}
Apart from the spectroscopic observations, we have also utilized the line-of-sight (LOS) photospheric magnetic field observations from the Helioseismic and Magnetic Imager (HMI; \cite{2012SoPh..275..207S}) and imaging observations (i.e., extreme ultraviolet (EUV) and H-$\alpha$ observations) from Atmospheric Imaging Assembly (AIA; \cite{lemen2012}) and the Global Oscillation Network Group (GONG; \cite{1996Sci...272.1284H}). The H-$\alpha$ observations are taken from the Big Bear Solar Observatory (BBSO) which is the part of the GONG program). In addition, we have also used the observations from the X-ray sensor (XRS) onboard Geostationary Operational Environment Satellites (GOES). 

\section{Observational Results}\label{sect:res}

\subsection{Overview of the Event}
The GOES measure the X-ray emission from the Sun in soft X-ray (i.e., 1.0{--}8.0~\AA) and hard X-ray (i.e., 0.5{--}4.0~\AA) channels. Based on the maximum soft X-ray (SXR) flux, the solar flares are classified into different classes: A-, B-, C-, M-, and X-class. Figure \ref{fig:fig_goes} shows GOES SXR (red curve) and hard X-ray (blue curve) variations from 14:00 to 20:00~UT. The gray area in Figure~\ref{fig:fig_goes} outlines the flare duration. The SXR started to rise at 16:24~UT, indicated by a black arrow. The SXR flux peaks approximately at a value of 7$\times$10$^{-7}$ W/m$^{2}$ at the 16:54~UT, i.e., at the flare's maximum phase which is indicated by the magenta arrow. Therefore, as per the flare classification, it is a B-class flare; see classification on the right side axis of Figure \ref{fig:fig_goes}. The SXR flux starts to decrease after 16:54~UT. Later, around 19:30~UT, the SXR flux reaches the same level as it was before the solar flare. It suggests that the solar flare is completed by 19:30~UT, indicated by an orange arrow in Figure~\ref{fig:fig_goes}. Opposite to SXR, the hard X-ray flux (blue curve) remains almost constant over this period of flare dynamics.

\begin{figure*}
\centering
\includegraphics[trim=0.0cm 0cm 0.0cm 0.0cm, scale=0.5]{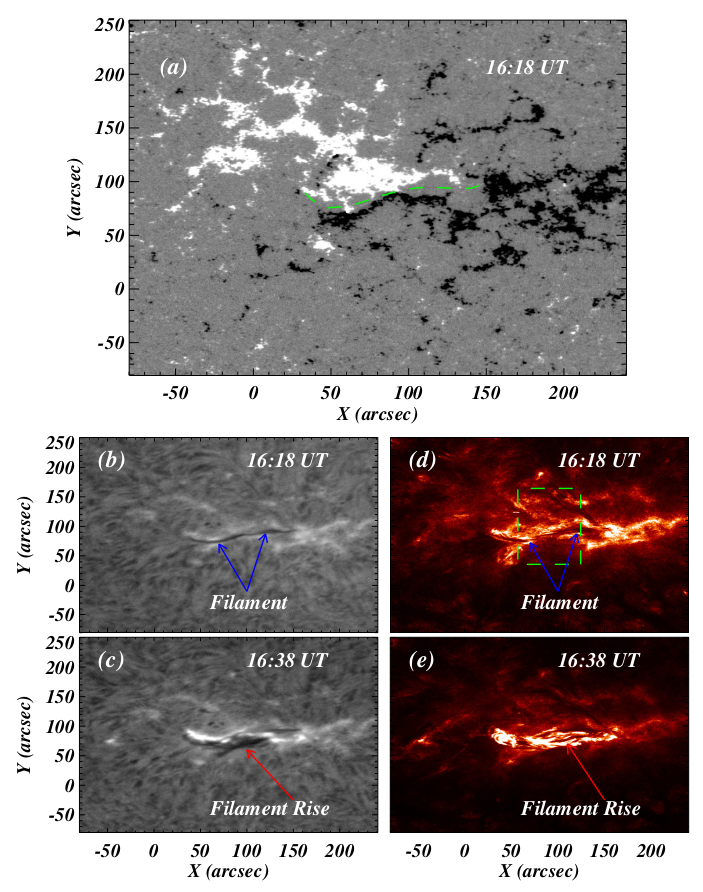}
\caption{The LOS magnetogram (panel a), H-$\alpha$ images (panels (b) and (c)), and AIA~304~{\AA} images (panels (d) and (e)) of the AR12661. The filament is visible in  H-$\alpha$ and AIA~304~{\AA} observations and the filament is indicated by the blue arrows (panels (b) and (d)). The green dashed line on panel (a) outlines the filament. The panels (c), and (e) show the filament raise from H-$\alpha$ and AIA~304~{\AA} observations, respectively. The IRIS raster FOV is outlined by the green dashed box on panel (d).}
\label{fig:fig_ref}
\end{figure*}

Panel (a) of Figure~\ref{fig:fig_ref} shows the LOS magnetic field before the flare, and the black (white) region corresponds to negative (positive) polarity. Next, panel(b) shows an H-$\alpha$ image before the flare, a dark thread-like structure, indicated by blue arrows, is a filament. The location of the filament is extracted from panel (b), and overplotted on the HMI magnetogram, see the green dashed line in panel(a). The same filament (as seen in panel b) is visible in AIA~304~{\AA} (panel d). Later on, the filament
rises which is visible in H-$\alpha$ and AIA-304, see the red arrows in panels (c) and (e). Finally, the filament erupts after 16:40~UT (not shown here).\\ 
The green dashed rectangular box in panel (d) shows the field of view, for which IRIS has captured the spectra. It is worth noting that filament rise/eruption overlaps with the lower part of the flare ribbon. As the downflow (in the flare ribbon) and upflow (in filament rise/eruption) motion mixes in the lower part of the flare ribbon, therefore, we see complex spectra even from optically thin Si~{\sc iv} and O~{\sc iv} lines, see section~\ref{sect:siv} and ~\ref{sect:mgii}     
\subsection{Si~{\sc iv} Analysis}\label{sect:siv}
\begin{figure}
\centering
\includegraphics[trim=4.0cm 0.0cm 0.0cm 0.0cm, scale=0.5]{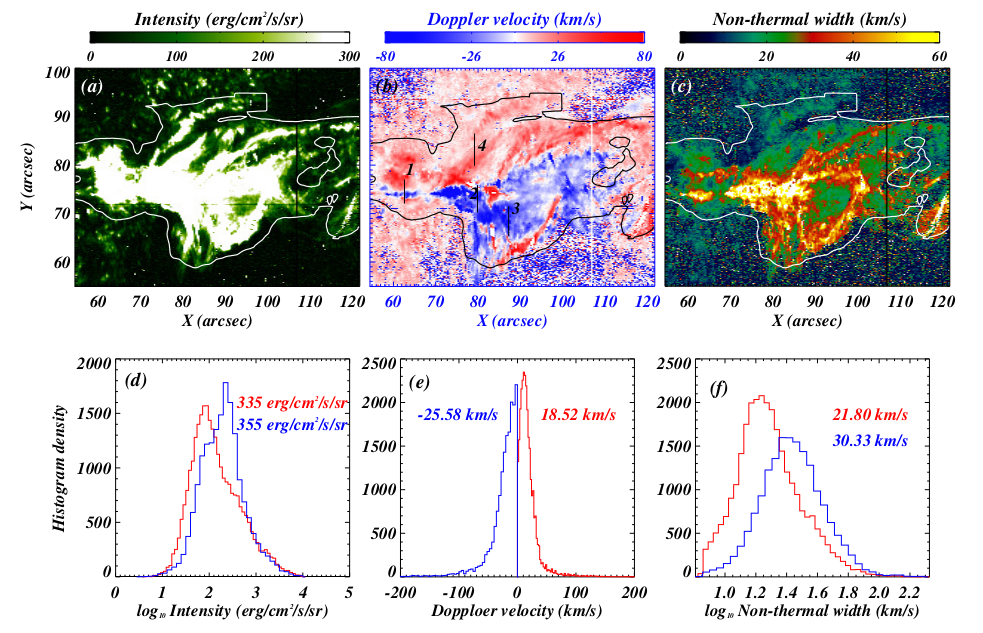}\\
\caption{The I$_{tot}$ (panel a), v$_{d}$ (panel b), and the v$_{nt}$ maps (panel c) deduced using the Si~{\sc iv} 1402.77~{\AA}. The contours drawn on each panel are based on the intensity threshold to outline the flare ribbons. Using v$_{d}$, we separated the blueshifted and the redshifted regions within the contour to plot the histogram. The panels (d), (e), and (f) show the I$_{tot}$, v$_{d}$, and v$_{nt}$ histograms for both the redshifted (red histogram) and blueshifted regions (blue histogram). The mean values of all the parameters are mentioned in each panel. Four slits are drawn on the v$_{d}$ map, and the key spectral profiles along these slits are shown in Figure~\ref{fig:si_spectra}.}
\label{fig:si_maps}
\end{figure}

The panels (a), (b), and (c) of Figure~\ref{fig:si_maps} show the I$_{tot}$, v$_{d}$, and v$_{nt}$ maps of the observed region from the first raster file. The observation time for this raster file is from 16:24~UT (flare initiation) to 16:40~UT. We see the flare ribbon (bright intensity area) in the I$_{tot}$ map (Figure~\ref{fig:si_maps}(a)). The intensity threshold of 50 erg/cm$^2$/s/sr is applied to locate the flare ribbons within the observed region. The white contour in panel (a) outlines the flare ribbons. The same contour is overplotted on the panels (b) and (c). There is one more file that covers the time after the first file i.e., from 16:40 to 16:56~UT.\\

The flare ribbons are redshifted in Si~{\sc iv} (Figure~\ref{fig:si_maps}(b)). However, not only the redshifts but blueshifts also exit inside the flare ribbons, see, the lower blueshifted region inside the black contour (Figure~\ref{fig:si_maps}(b)). This blueshifted region is the overlapping region of filament rise/eruption and flare ribbon. Please note that in this event, the filament rises just after the flare initiation, later, the filament erupts (see \ref{sect:res}). The lower blueshifted region within the flare ribbon is due to the filament rise/eruption (see Figure~\ref{fig:fig_ref}, \ref{sect:siv}). Further, from the flare ribbons of raster files 1 and 2, we have gathered line parameters (i.e., I$_{tot}$, v$_{d}$, and v$_{nt}$) corresponding to redshift and blueshift regions, separately, and produced histograms for I$_{tot}$ (Figure~\ref{fig:si_maps}(d)), v$_{d}$ (Figure~\ref{fig:si_maps}(e)), and v$_{nt}$ (Figure~\ref{fig:si_maps}(f). The red (blue) histogram in panels (d), (e), and (f) of Figure~\ref{fig:si_maps} corresponds to the redshifted (blueshifted) regions. The mean of I$_{tot}$, v$_{d}$, and v$_{nt}$ are 335$\pm$683.02 erg/cm$^2$/s/sr, 18.52$\pm$18.75 km/s, and 21.80$\pm$12.70 km/s in the redshifted region. While, the mean of I$_{tot}$, v$_{d}$, and v$_{nt}$ in the blueshifted regions are 335$\pm$604.31 erg/cm$^2$/s/sr, -25.58$\pm$31.04 km/s, and 30.33$\pm$15.83 km/s. The I$_{tot}$ is almost the same in both regions while v$_{nt}$ is significantly higher in the blueshifted region than in the redshifted region. Further, on average, the blueshift is stronger than the redshift.\\

\begin{figure}
\centering
\includegraphics[trim=2.8cm 0.0cm 0.0cm 0.0cm, scale=0.4]{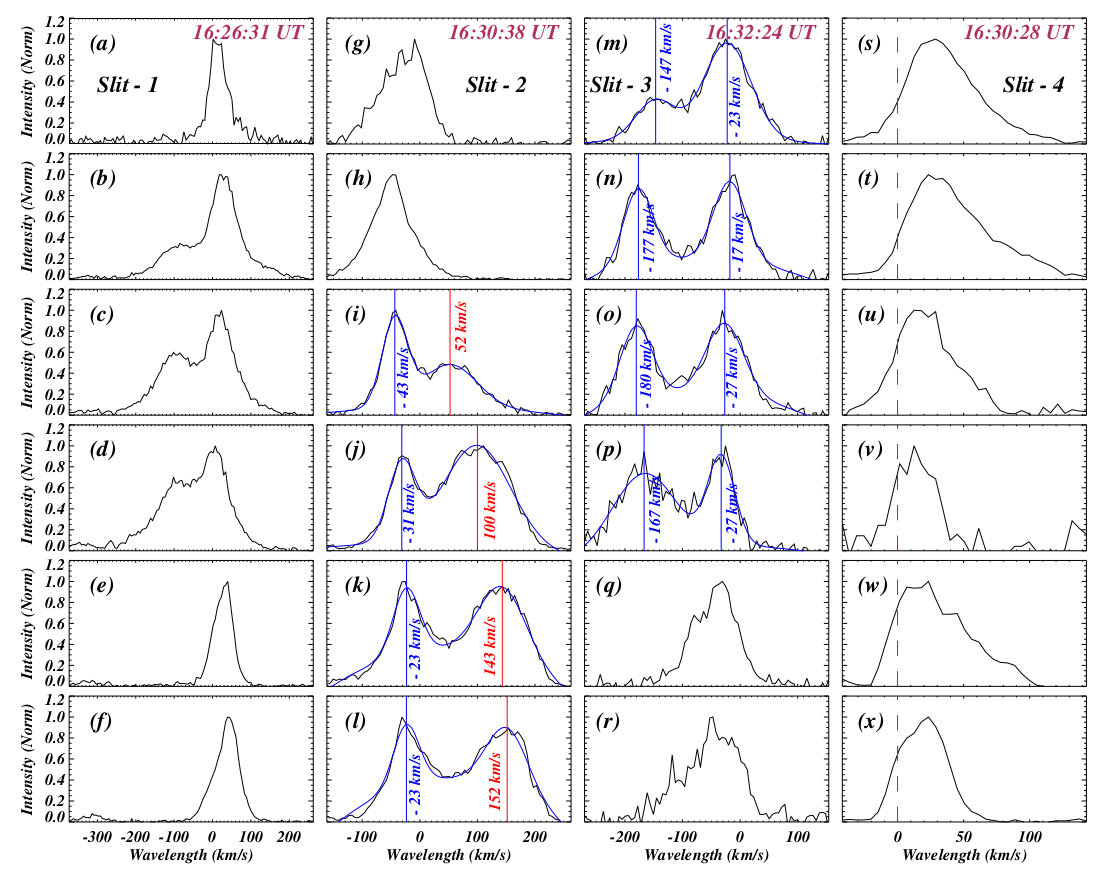}\\
\caption{The key spectral profiles along the slit 1 (panels (a) to (f)), slit 2 (panels (g) to (l)), slit 3 (panels (m) to (r)), and slit 4 (panels (s) to (x)). All four slits are displayed in the panel (b) of figure~\ref{fig:si_maps}. The overplotted blue curve in some panels is the double Gaussian fit on the observed profile. The position of each peak is shown by blue/red solid vertical lines in the corresponding spectra, and the v$_{d}$ of each component is mentioned in those panels. The vertical red dashed line in the last column (panels (x) to (s)) is at rest wavelength.}
\label{fig:si_spectra}
\end{figure}
The Si~{\sc iv} spectral profiles are investigated along four different slits drawn in Figure~\ref{fig:si_maps}(b)). The slit 1 (observed at 16:26:31 UT) crosses the small blueshifted region within the redshifted region. Some of the key profiles along slit 1 are displayed in Figure~\ref{fig:si_spectra}, see panels (a) to (f). The spectral profiles from the blueshift region are double peak profiles (panels (b), (c), and (d)), and the redshift region has single peak profiles only (panels (a), (e), and (f)).\\

Next, slit 2 (observed at 16:30:38 UT) lies in the overlapping region, and the key profiles from this slit are displayed in panels (g) to (l). The majority of the spectral profiles (i.e., more than 65\%) along this slit have double peaks (see panels from (i) to (l)). Interestingly, one peak is in the redshift (vertical red lines in panels (i) to (l)) while the other peak is in the blueshift (vertical blue line in panels (i) to (l)). It would be worth noting that redshift reaches more than 150 km/s (panel (l)). Meanwhile, the maximum blueshift is around 43 km/s (panel (i)). The slit 3 is also located in the overlapping region but observed at 16:32:24 UT, and the key profiles are shown in panels (m) to (r). Similar to slit 2, more than 65$\%$ profiles are double peak profiles. Although, compared to slit 2, the blueshift in slit 3 has increased up to 180 km/s (panels (n) and (o)). While, the other peak, which was in redshift in slit 2 (panels (j), (k), and (l)), is now shifted to the blueshifts of more than 30 km/s (panel (p)). Lastly, we mention that not only along slits 2 and 3, but the majority of the profiles within the overlapping region are also double peak profiles.\\ 

Finally, slit 4 is located in the redshift region only (flare ribbon), and it was observed at 16:30:28 UT. Some key spectral profiles from the slit 4 are displayed in panels (s) to (x). All the spectral profiles are single peak profiles with the dominance of redward asymmetry. The vertical red-dashed line in the panels (s) to (x) is at rest wavelength. Again, we mention that not only along slit 4, but the majority of the profiles within the redshifted region are single peak profiles with the dominance of the redward asymmetry.\\

\subsection{O~{\sc iv} Analysis}\label{sect:oiv}
\begin{figure}
\centering
\includegraphics[trim=3.0cm 0.0cm 0.0cm 0.0cm, scale=0.3]{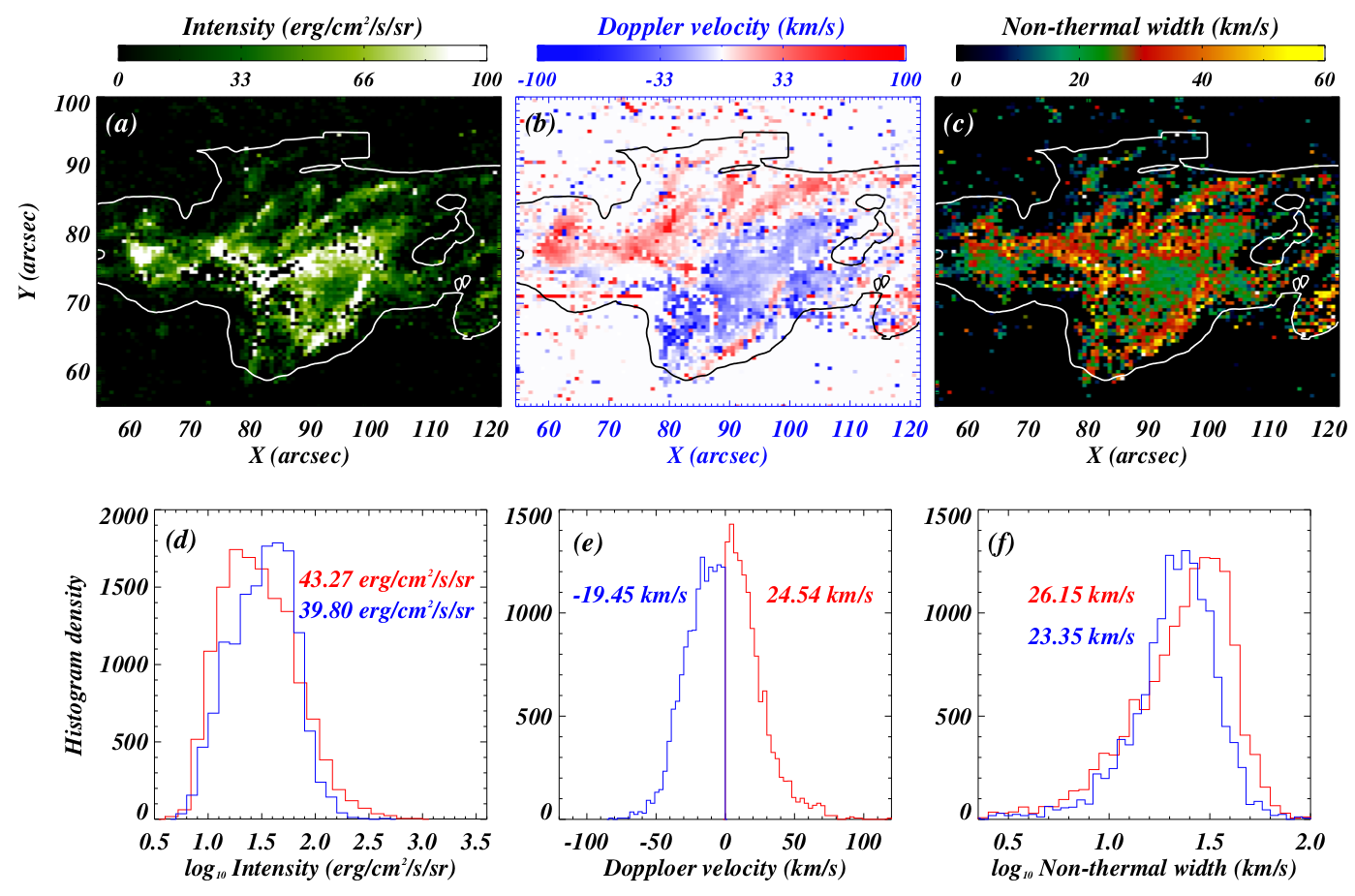}\\
\caption{The same as figure~\ref{fig:si_maps} but using the O~{\sc iv} 1401.16~{\AA}.}
\label{fig:o_maps}
\end{figure}

The I$_{tot}$, v$_{d}$, and v$_{nt}$ maps from O~{\sc iv}~1401.16~{\AA} line are shown in panels (a), (b), and (c) of Figure~\ref{fig:o_maps}. We have overplotted the same intensity contour (\ref{sect:siv}) in the panels (a), (b), and (c) of Figure~\ref{fig:o_maps}. 
Similar to Si~{\sc iv}, the redshifted and blueshifted regions exist within the contour, and further we collected line parameters from the redshifted and blueshifted regions separately. Finally, the histograms of I$_{tot}$, v$_{d}$, and v$_{nt}$ are displayed in the panels (d), (e), and (f) of Figure~\ref{fig:o_maps}.
The mean I$_{tot}$ from the redshifted and blueshifted regions are 43.27$\pm$52.39 erg/cm$^2$/s/sr and 39.80$\pm$28.32 erg/cm$^2$/s/sr, respectively. The mean v$_{d}$ are -19.45$\pm$13.49 km/s and 24.54$\pm$32.87 km/s from blueshifted and redshifted regions of flare-ribbon, respectively. The mean v$_{nt}$ from the redshifted region is 23.35$\pm$10.39 km/s and from the blueshifted regions is 26.15$\pm$12.95 km/s.    

\subsection{Mg~{\sc ii} k Analysis}\label{sect:mgii}
\begin{figure}
\centering
\includegraphics[trim=2.5cm -0.5cm 0.0cm 0.0cm, scale=0.3]{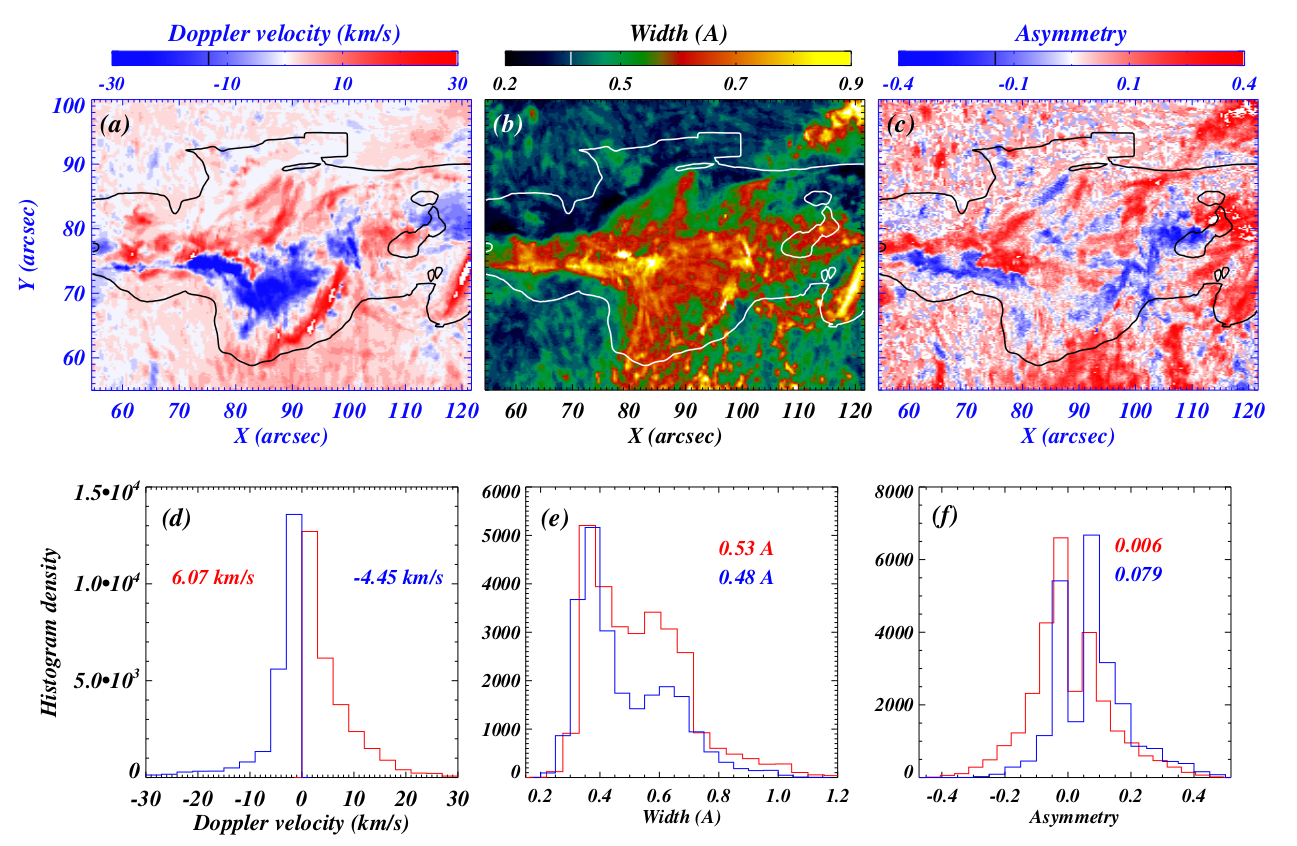}\\
\caption{The figure shows the v$_{d}$ (panel a), line width (panel b), and line asymmetry (panel c) maps deduced using Mg~{\sc ii} k 2796.35~{\AA}. The same contour, as used to draw the contour on the Si~{\sc iv} maps and O~{\sc iv} maps (Figure~\ref{fig:si_maps} and ~\ref{fig:o_maps}), is drawn in the panels (a), (b), and (c). The v$_{d}$, line width, and asymmetry histograms are shown in panels (d), (e), and (f), respectively. In these panels, the red histogram corresponds to the flare ribbon, and the blue histogram corresponds to the filament rise/eruption.}
\label{fig:mg_maps}
\end{figure}

The v$_{d}$ (panel a), line width (panel b), and asymmetry maps (panel c) from Mg~{\sc ii} k~2796.35~{\AA} are displayed in Figure~\ref{fig:mg_maps}. 
Similar to the Si~{\sc iv} and O~{\sc iv}, the redshifted and blueshifted patches exist in Mg~{\sc ii} k line (panel a; Figure~\ref{fig:mg_maps}). Again, we gathered v$_{d}$, line width, and asymmetry from the blueshifted and redshifted regions. Then the histogram of v$_{d}$, line width, and asymmetry are displayed in panels (d), (e), and (f) of Figure~\ref{fig:mg_maps}. The red and blue histograms in each panel are from redshifted and blueshifted regions, respectively. The mean Doppler velocities are 6.07$\pm$5.63 km/s and -4.45$\pm$7.52 km/s in the redshifted and blueshifted regions, respectively (panel d). The mean line width is almost the same in both regions (panel e). But, more interestingly, the line width reaches up to 1.2~{\AA}. The mean asymmetries peak around zero in both regions (panel f), although the asymmetry varies from -0.4 to +0.4 in the redshifted region and -0.2 to +0.4 in the blueshifted region.
\begin{figure}
\centering
\includegraphics[trim=2.5cm 0.0cm 0.0cm 0.0cm, scale=0.3]{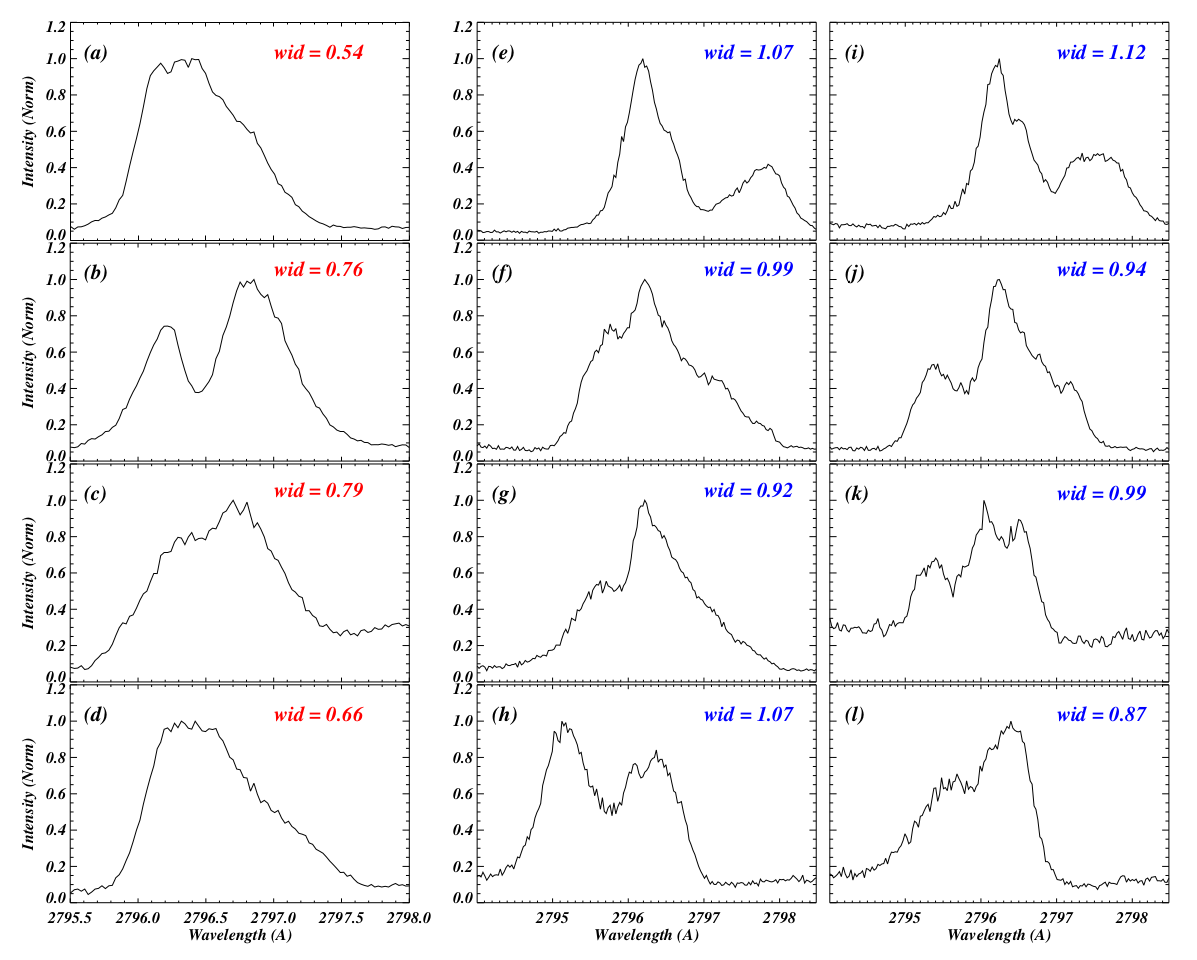}\\
\caption{The figure shows some important spectral profiles of Mg~{\sc ii} k 2796.35~{\AA} with different spectral widths. The first column (panels (a) to (d)) shows the profiles from the flare ribbon. The middle and right columns show the profiles from the blueshifted region. The width of each spectrum is mentioned in the respective panels.}
\label{fig:mg_spectra}
\end{figure}
Finally,  Figure~\ref{fig:mg_spectra} shows some Mg~{\sc ii} k spectral profiles with high line widths. The left column (panels (a) to (d)) shows the profiles from the redshifted region while the middle and right columns (panels (e) to (l)) show the spectral profiles from the blueshifted region. The line width is written in each panel. 
\section{Summary and Discussion}\label{sect:sec_sum}
The main findings of the work are summarized below.
\begin{itemize}
\item The flare-ribbons are redshifted in O~{\sc iv} 1401.16~{\AA}, Si~{\sc iv}~1402.77~{\AA}, and Mg~{\sc ii} k~2796.35~{\AA} lines. The mean redshift decreases with the decreasing temperature, i.e., 24.54 km/s at log T = 5.2 (O~{\sc iv}), 18.52 km/s at log T/K = 4.8 (Si~{\sc iv}), and 6.07 km/s at log T/K = 4.0 (Mg~{\sc ii}).

\item The filament rises/erupts in the vicinity of the lower part of the flare ribbons i.e., overlapping region). Therefore, the lower part is dominated by the blueshifts. The mean blue-shifts are -19.45 km/s in O~{\sc iv}, -25.58 km/s in Si~{\sc iv}, and -4.45 km/s in Mg~{\sc ii} k.

\item The mean v$_{nt}$ of the flare ribbon are higher in the O~{\sc iv} than Si~{\sc iv}. On the other hand, the mean v$_{nt}$ in the overlapping region is lower in O~{\sc iv} than Si~{\sc iv}. There is no significant difference between the Mg~{\sc ii} line widths of the overlapping region and flare ribbon. But, most importantly, the line width of Mg~{\sc ii} can reach up to 1.20~{\AA}.

\item The Si~{\sc iv} spectral profiles from the flare ribbon are single peak profiles with significant red asymmetry. While most of the Si~{\sc iv} profiles from the overlapping region are double peak profiles. Although, the O~{\sc iv} line is weak in comparison to the Si~{\sc iv} but the spectral behaviour is similar to the Si~{\sc iv} line. 

\item The Mg~{\sc iv} line shows significant asymmetry, varying from $-$0.4 to $+$0.4.

\end{itemize}

Generally, the flare ribbons in Si~{\sc iv} do exhibit redshift less than 50 km/s, and, in the energetic M- or X-class flares, the redshift can increase up to 100 km/s (e.g. \cite{2014ApJ...797L..14T, 2015ApJ...811..139T, 2016ApJ...832...65Z, 2020ApJ...904...95Z}).  Particularly, the redshifts, in weak B-class flare/microflares, vary from 2 km/s to 40 km/s within the chromosphere/TR (e.g., \cite{2009A&A...505..811B, 2013A&A...550A..16G, 2019ApJ...881..109H, 2023FrASS..1014901P}). The present work also shows a similar range of mean redshifts in the chromosphere/TR, i.e., and 
6.07 km/s in Mg~{\sc ii}, 18.52 km/s in Si~{\sc iv}, and 24.54 km/s in O~{\sc iv}. Most interestingly, in the present observation, we found that Si~{\sc iv} redshifts have increased beyond 150 km/s (see panels (k) and (l); Figure~\ref{fig:si_spectra}). Here, it should be noted that \cite{2009A&A...505..811B} has also shown the existence of very strong downflow in a microflare (i.e., redshift of around 180 km/s in He~{\sc i} and O~{\sc v}) but in very few locations. 
Recently, \cite{2023ApJ...944..104P} showed that flare-ribbons in O~{\sc iv} line dominate by the redshift but some locations within the flare ribbons have blueshifts also. In addition, \cite{2023ApJ...944..104P} have also shown that the O~{\sc iv} v$_{d}$ varies from -40 to 40 km/s. The present work shows redshifts of O~{\sc iv} reach up to 50 km/s. Unlike \cite{2023ApJ...944..104P}, within the flare ribbon, we couldn't find the blueshifts in O~{\sc iv}. The disappearance of blueshifts in flare ribbons might be due to the weak strength of the flare. But, the blueshifts in O~{\sc iv} due to filament rise/eruption reach up to -50 km/s. In the case of the chromosphere, the v$_{d}$ varies from 15 to 30 km/s in M-class flare as reported by \cite{kerr2015, 2015ApJ...807L..22G, 2020ApJ...895....6G}. The current works show that mean Mg~{\sc ii} v$_{d}$ in the flare ribbons is around 6.07 km/s but it does reach up to 30 km/s. Further, the mean blueshift due to filament eruption is around 5 km/s but in some locations,v it can reach up to -20 km/s.\\


The present observation shows significant number of double peak Si~{\sc iv} and O~{\sc iv} profiles in the overlapping region, i.e., a region in which filament rise/eruption and flare ribbons overlap. On-average, in the overlapping region, the spectral profiles are blueshifted. Of course, the overlapping region belongs to filament rise/eruption but, at the same time, the chromospheric condensation due to solar flare is also operating in the same region. Additionally, the filament material also drains (i.e., downflow of the filament plasma) during its rise phase (e.g., \cite{2018SoPh..293....7J, 2021ApJ...923...74D}). The upflow of plasma is due to the filament rise/eruption, while the downflow is due to chromospheric condensation and/or filament mass drainage. So, the upflows and downflows exist simultaneously in the overlapping region. Note that the simultaneous existence of upflow and downflow can lead to the double peak profiles. Previously, in sunspot umbra, the double peak spectral lines, which originate from TR/chromosphere (i.e., Si~{\sc iv}, C~{\sc ii}, and Mg~{\sc ii}), are very frequent due to simultaneous existence of upflows (due to passage of shocks) and downflows (due to plasma falling back from previous shock) in the same region (e.g., \cite{2006ApJ...640.1153C, 2014ApJ...786..137T, 2021ApJ...906..121K}). Therefore, one peak is in the redshift while another peak shows the blueshift. Interestingly, in the initial phase, when the filament rises slowly, one peak is in the redshift (plasma downflow due to chromospheric condensation and/or filament mass drainage) and the other peak is in the blue shift (plasma upflow due to filament rise), see the panels (b), (c), (d), (i), (j), (k), and (l) of Figure~\ref{fig:si_spectra}. However, without the involvement of filament rise/eruption, the existence of two or sometimes three peaks in microflares has already been reported (e.g., \cite{2009A&A...505..811B, 2013A&A...550A..16G}).\\

As mentioned above, the downflow (redshift) can be either due to chromospheric condensation or filament mass drainage. Also, the downflow motion can be the result of both physical processes occurring simultaneously. In the flare ribbons, only the chromospheric condensation is operating. While, in the overlapping region, filament drainage and chromospheric condensation might operate simultaneously. The redshift of more than 150 km/s is found in the overlapping region (see profiles in the second and third columns of Figure~\ref{fig:si_spectra}). In contrast, the redshift is just around 40 km/s in the flare ribbons (profiles in the last column~\ref{fig:si_spectra}). So, the redshift is very high in overlapping regions compared to the flare ribbons. Most probably, the high redshift in the overlapping region is a result of the simultaneous existence of filament drainage and chromospheric condensation. However, in the present observational baseline, we can't distinguish whether the redshift is due to chromospheric condensations or filament mass drainage.\\

But at the later time (t = 16:32:24 UT; slit 3), some profiles have both peaks into the blueshifts (upflows) only, panels (m), (n), (o), and (p) of Figure~\ref{fig:si_spectra}. One peak has an extremely high blueshift (i.e., around -180 km/s; panels n and o) while another peak has a weak upflow velocity of around -30 km/s (panel l). The filament's material is non uniformly distributed, i.e., several different regions (say plasma threads) within the filament. These regions can have different densities and different upflow velocities. Most probably, both blueshifted peaks might result from the superposition of two different velocity distributions (i.e., strong upflow and weak upflow) coming from two distinct regions within the filament in the line of sight. Later on, at the same time, but from a different location, the single peak profiles (see panels (q) and (r); Figure~\ref{fig:si_spectra}) exit which have blueshift. As already stated filament material is non-uniformly distributed, therefore, it might be possible that the filament material with strong upflow (i.e., strong velocity distribution) does not exist in this area. And, only filament material with weak upflow velocity exists in this region. Therefore, this region is affected by the filament material having a relatively slow upflow speed, thus, the profiles, from the lower edge of slit 3 (i.e., panels (q) and (r)), are single peak profiles with weak blueshift. In the flare ribbons, all the profiles are single peaks with redshifts of around 25 to 35 km/s (right-most column; Figure~\ref{fig:si_spectra}). Most importantly, apart from the redshift, all the profiles also have redward asymmetry in the flare ribbon. This redshift/redward asymmetry is the result of chromosphere condensations (e.g., \cite{Fletcher2011, 2014ApJ...786..137T, 2017ApJ...846....9K, K_Yu2020}).\\

In the B-class flare, the v$_{nt}$ varies 30{--}45 km/s in narrow component of C~{\sc iv} line (\cite{2013A&A...550A..16G}). While, the v$_{nt}$ from broader component of C~{\sc iv} was extremely high (i.e., 80{--}140 km/s; \cite{2013A&A...550A..16G}). Further, \cite{2018SciA....4.2794J} showed that v$_{nt}$ from Si~{\sc iv} line reached up to only 30 km/s in a B-class flare ribbon. 
In the present B-class flare, the mean v$_{nt}$ in flare ribbons is around 22 km/s in Si~{\sc iv}. While on the higher side, the v$_{nt}$ has reached up to 80 km/s at some locations. Compared to Si~{\sc iv}, the v$_{nt}$ of the flare ribbons is higher in O~{\sc iv} line. On the contrary, v$_{nt}$ of O~{\sc iv} from the overlapping region is weaker than Si~{\sc iv}. In Mg~{\sc ii}, the v$_{nt}$ is not estimated but we estimate the line width. The line width of Mg~{\sc ii} k~2796.35~{\AA} should be around 0.02{--}0.04~{\AA} for the typical chromospheric temperature (\cite{kerr2015}). The flare ribbons heat the chromosphere significantly, and the line width increases of the Mg~{\sc ii} line. Although, it is implausible that the increased width is due to the heated temperature (\cite{kerr2015}). In an M-class flare, \cite{kerr2015} reported that line width can increase from 0.28~{\AA} (pre-flare conditions) to 0.55~{\AA} (flare maximum). Interestingly, the line width is very high in the present B-class solar flare. At various locations, the line width is higher than 0.70~{\AA} (Figure~\ref{fig:mg_spectra}). Lastly, in some cases, the line width has increased beyond 1.10~{\AA} (panel (i); Figure~\ref{fig:mg_spectra}) which is an extremely high value.\\


The Mg~{\sc ii} line asymmetry shows a mixture of reward and blueward asymmetry in the flare ribbon and overlapping region. The reward asymmetry is due to the chromospheric condensation (e.g., \cite{Fletcher2011, 2015ApJ...813..125K}). While the blueward asymmetry locations are the regions of the strongest emission, and in these regions, upflow and downflow exist (\cite{kerr2015}). Another interpretation of blueward asymmetry exists in the chromosphere is given \cite{1994SoPh..152..393H}; the downward propagating plasma can absorb the radiation from the redward side of the line core, and this weakens the redward side and the blueward side becomes strong, i.e., origin of the blueward asymmetry. However, in the current observational baseline, we cannot confirm which process is responsible for blueward asymmetry in the Mg~{\sc ii} line.\\

\bmhead{Acknowledgements}
We thank the anonymous referee for his/her valuable comments/suggestions. IRIS is a NASA small explorer mission developed and operated by LMSAL with mission operations executed at NASA Ames Research Center and major contributions to downlink communications funded by ESA and the Norwegian Space Centre. We also acknowledge the use of SDO/AIA and SDO/HMI observations for this study. 

\section*{Declarations}

\begin{itemize}
\item Funding - The authors declare that no funds, grants, or other supports were received for this research work.
\item Conflict of interest/Competing interests - The authors declare no conflict/competing interests.
\item Ethics approval and consent to participate - Not Applicable
\item Consent for publication - Not Applicable
\item Data availability - The data underlying this article are available at \url{https://iris.lmsal.com/data.html} (NASA/IRISwebsite) and at \url{https://iris.lmsal.com/search/}(LMSAL search website). Note that IRIS data are publicly available with the observation ID OBS 3880356995. A direct link to the data used in this manuscript is provided below.\\
\url{https://www.lmsal.com/hek/hcr?cmd=view-event&event-id=ivo%3A%2F%2Fsot.lmsal.com%2FVOEvent%23VOEvent_IRIS_20170607_135931_3641257172_2017-06-07T13%3A59%3A312017-06-07T13%3A59%3A31.xml}
\item Materials availability - Not Applicable
\item Code availability - Not Applicable
\item Author contribution - The data collection and analysis were done by
BSB with suggestions provided by PK. BSB has written the first
draft of the manuscript. PK made the final draft of
the manuscript. SCT contributed to some language corrections in the
manuscript.
\end{itemize}



\end{document}